\documentclass[sigconf]{acmart}

\usepackage{makecell}
\usepackage{subcaption}
\usepackage{graphicx}

\AtBeginDocument{%
  }

\setcopyright{acmlicensed}
\copyrightyear{2026}
\acmYear{2026}
\acmDOI{XXXXXXX.XXXXXXX}
\acmConference[SIGIR '2026]{The 49th International ACM SIGIR Conference}{20–24 July 2026}{Melbourne | Naarm, Australia}

\acmSubmissionID{681}

\begin{document}

\title{KARMA: Knowledge-Action Regularized Multimodal Alignment for Personalized Search at Taobao}


\author{Zhi Sun}
\affiliation{%
  \institution{Taobao $\&$ Tmall Group of Alibaba}
  \city{Hangzhou}
  \country{China}
}
\email{sunzhi.sun@taobao.com}

\author{Wenming Zhang}
\affiliation{%
  \institution{Taobao $\&$ Tmall Group of Alibaba}
  \city{Hangzhou}
  \country{China}
}
\email{zhangwenming.zwm@taobao.com}

\author{Yi Wei}
\affiliation{%
 \institution{Taobao $\&$ Tmall Group of Alibaba}
  \city{Hangzhou}
  \country{China}
}
\email{songwen.wy@alibaba-inc.com}

\author{Liren Yu}
\affiliation{%
  \institution{Taobao $\&$ Tmall Group of Alibaba}
  \city{Hangzhou}
  \country{China}
}
\email{yuliren.ylr@taobao.com}

\author{Zhixuan Zhang}
\affiliation{%
  \institution{Taobao $\&$ Tmall Group of Alibaba}
  \city{Hangzhou}
  \country{China}
}
\email{zhibing.zzx@taobao.com}

\author{Dan Ou}
\affiliation{%
  \institution{Taobao $\&$ Tmall Group of Alibaba}
  \city{Hangzhou}
  \country{China}
}
\email{oudan.od@taobao.com}

\author{Haihong Tang}
\affiliation{%
  \institution{Taobao $\&$ Tmall Group of Alibaba}
  \city{Hangzhou}
  \country{China}
}
\email{piaoxue@taobao.com}

\renewcommand{\shortauthors}{Trovato et al.}

\begin{abstract}
Large Language Models (LLMs) are equipped with profound semantic knowledge, making them a natural choice for injecting semantic generalization into personalized search systems. However, in practice we find that directly fine-tuning LLMs on industrial personalized tasks (e.g. next item prediction) often yields suboptimal results. We attribute this bottleneck to a critical Knowledge--Action Gap: the inherent conflict between preserving pre-trained semantic knowledge and aligning with specific personalized actions by discriminative objectives. Empirically, action-only training objectives induce Semantic Collapse, such as attention ``sinks''. This degradation severely cripples the LLM's generalization, failing to bring improvements to personalized search systems.

We propose KARMA (Knowledge--Action Regularized Multimodal Alignment), a unified framework that treats semantic reconstruction as a train-only regularizer. KARMA optimizes a next-interest embedding for retrieval (Action) while enforcing semantic decodability (Knowledge) through two complementary objectives: (i) history-conditioned semantic generation, which anchors optimization to the LLM's native next-token distribution, and (ii) embedding-conditioned semantic reconstruction, which constrains the interest embedding to remain semantically recoverable. 

On Taobao search system, KARMA mitigates semantic collapse (attention-sink analysis) and improves both action metrics and semantic fidelity. In ablations, semantic decodability yields up to +22.5 HR@200. With KARMA, we achieve +0.25 CTR AUC in ranking, +1.86 HR in pre-ranking and +2.51 HR in recalling. Deployed online with low inference overhead at ranking \& pre-ranking stage, KARMA drives +0.9\% increase in GMV.
\end{abstract}



\maketitle

\section{Introduction}

In personalized search systems, ranking models are predominantly optimized using post-hoc user feedback(e.g. CTR prediction)\cite{covington2016deep,cheng2016wide,zhou2018deep,yu2025hhft}. This optimization paradigm inevitably privileges post-hoc memorization features (item IDs, co-occurrence statistics, etc. ) over semantic generalization features. While memorization features precisely capture user preferences for popular items, yielding excellent performance on head traffic, they expose a critical vulnerability: when faced with sparse feedback scenarios, such as long-tail queries or cold-start items, this lack of semantic generalization leads to a sharp decline in ranking quality. This naturally motivates leveraging the rich, pre-trained semantic knowledge of LLMs to compensate for the system's generalization capabilities\cite{zhang2024notellm,lee2024star,sheng2024enhancing,jia2025learn}.

Although integrating LLMs is intuitively appealing, directly fine-tuning LLMs for personalized ranking often yields limited performance gains. We attribute this empirical bottleneck to a critical Knowledge-Action Gap: the inherent conflict between rich pre-trained semantics (Knowledge) and user behavioral patterns driven by discriminative feedback (Action). This gap creates a severe trade-off: over-optimizing for Action severely distorts the pre-trained semantic space, while freezing the LLM yields representations too coarse-grained for personalization.

Constrained by latency, industrial systems typically compress historical items (text and image) into single embeddings\cite{chen2024hllm,yang2025sparse} for the next-item prediction task\cite{kang2018self}. However, this embedding-only adaptation triggers Semantic Collapse. We observe that attention maps degenerate into ``barcode-like'' patterns, acting as a destructive attention sink. By exploiting shortcut cues to distinguish positives from negatives, the LLM bypasses genuine semantic modeling, leading to severe semantic degradation.

\begin{figure}[htbp] 
    \centering 
    \includegraphics[width=\linewidth]{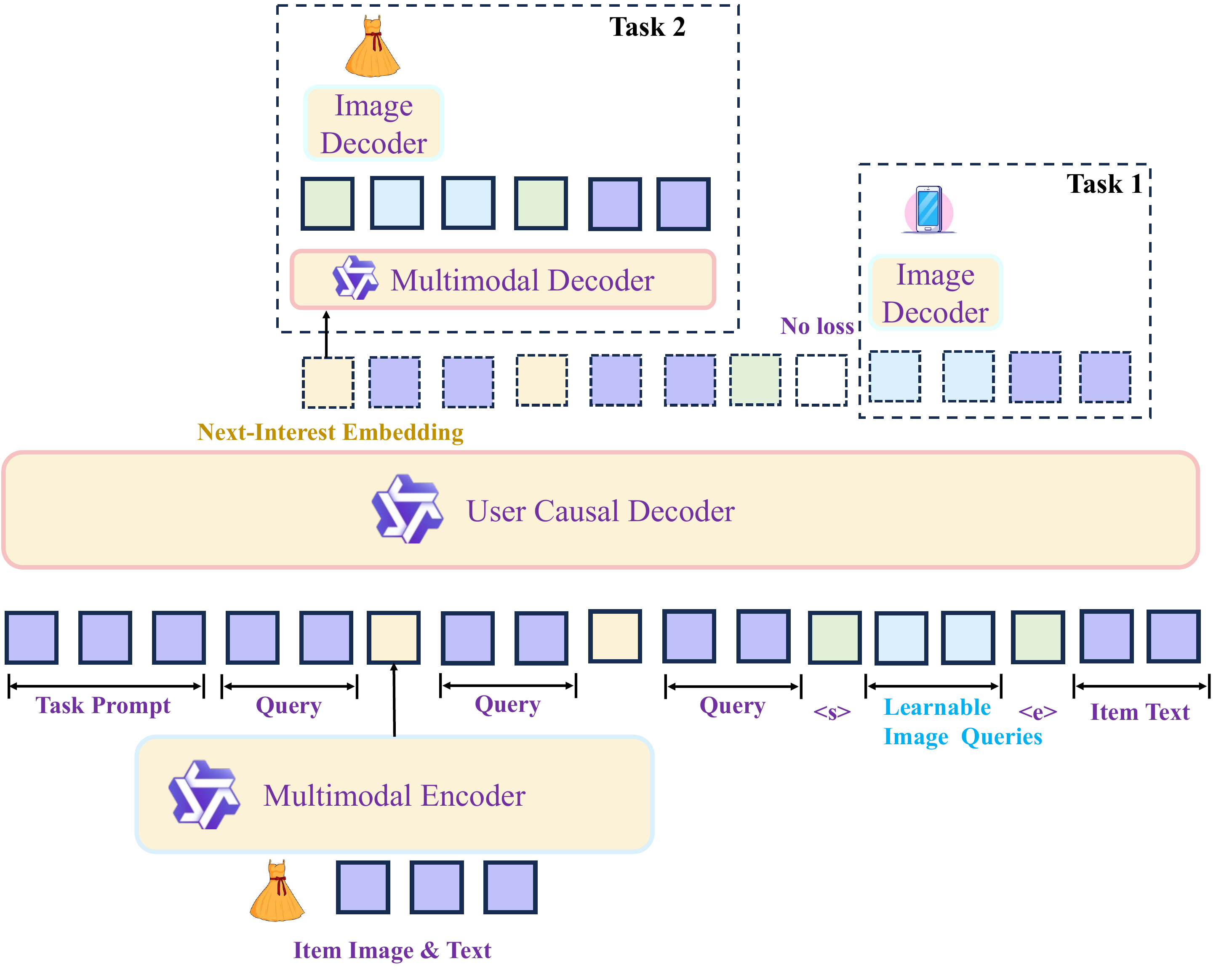} 
    \caption{KARMA architecture: A train-only regularizer that bridges the Knowledge--Action Gap.} 
    \label{fig:karma}
\end{figure}

This paper proposes a principle for continuous-token personalization: semantic decodability should be enforced as a train-only regularizer, rather than treated as an auxiliary generation capability. We introduce KARMA (Knowledge--Action Regularized Multimodal Alignment), which learns a next-interest embedding optimized for retrieval and ranking (Action), while preserving LLM semantics (Knowledge) via two non-conflicting decodability objectives. The first decodes the target item's semantics from the behavioral history, anchoring training to the LLM's native token distribution. The second decodes the same semantics conditioned explicitly on the model-produced next-interest embedding, enforcing an embedding bottleneck that prevents ID-like shortcut solutions. When multimodal supervision is available, we further reconstruct frozen visual semantic features as an additional grounding signal. Figure~\ref{fig:karma} shows the architecture of KARMA.

\textbf{Contributions.} (1) We identify Semantic Collapse as a key failure mode of LLM personalization under continuous-token interfaces, supported by attention-sink analysis and semantic metrics. (2) We propose KARMA, which bridges the Knowledge--Action Gap by enforcing semantic decodability as a train-only regularizer for action-aligned retrieval embeddings. (3) We demonstrate substantial offline and online gains in a large-scale search engine with zero inference overhead.

\section{Methodology}

\subsection{Continuous-Token Personalized Search}

At impression time $t$, user $u$ issues query $q_t$ with a chronological sequence of positive interactions
\begin{equation}
\mathcal{H}_t=\{(q_1,i_1),\ldots,(q_{t-1},i_{t-1})\}
\end{equation}
where each item $i$ contains multimodal content $\mathbf{x}_i=(\mathbf{x}_i^{\text{text}},\mathbf{x}_i^{\text{img}})$.
Our goal is to compute a dense next-interest representation $\mathbf{h}_t\in\mathbb{R}^d$ for retrieval and ranking.

\paragraph{Efficiency constraint.}
Feeding raw tokens of all historical items into an LLM is infeasible in industrial search. We therefore adopt a continuous-token interface: an item encoder compresses each item into a single embedding token
\begin{equation}
\mathbf{e}_i = E_\phi(\mathbf{x}_i)\in\mathbb{R}^d
\end{equation}
and a decoder-only Transformer $D_\theta$ performs causal modeling over the history token sequence $[\mathbf{e}_{i_1},\ldots,\mathbf{e}_{i_{t-1}}]$ (optionally conditioned on the current query via a lightweight query embedding). The final hidden state is used as the next-interest vector:
\begin{equation}
\mathbf{S}_{t-1}=D_\theta([\mathbf{e}_{i_1},\ldots,\mathbf{e}_{i_{t-1}}]),\qquad
\mathbf{h}_t=\mathbf{S}_{t-1}[-1]
\end{equation}
Item embeddings $\mathbf{e}_i$ are indexed for ANN retrieval and used for dot-product scoring.

This interface is computationally attractive, but it is also where we observe Semantic Collapse: when trained with purely discriminative objectives, the model may treat $\mathbf{e}_i$ as opaque identifiers and converge to shortcut, ID-centric solutions.

\subsection{KARMA: Knowledge--Action Regularization via Decodability}

KARMA is built on one requirement: the representation $\mathbf{h}_t$ should be action-aligned for ranking, while remaining semantically decodable back to the target item's original semantics. We optimize the following objective:
\begin{equation}
\mathcal{L}_{\text{KARMA}}
=
\mathcal{L}_{\text{act}}
+
\lambda_{\text{dec}}\mathcal{L}_{\text{dec}}
\end{equation}
where $\mathcal{L}_{\text{act}}$ injects behavioral supervision (Action), and $\mathcal{L}_{\text{dec}}$ regularizes the solution space by enforcing decodability (Knowledge).

\paragraph{Two decodability paths.}
We implement $\mathcal{L}_{\text{dec}}$ with two complementary paths.
(i) History-conditioned semantic generation decodes the target semantics directly from the behavioral history, anchoring optimization to the LLM's native next-token distribution.
(ii) Embedding-conditioned semantic reconstruction decodes the same semantics conditioned explicitly on the model-produced next-interest embedding $\mathbf{h}_t$, enforcing $\mathbf{h}_t$ to remain semantically recoverable and preventing ID-like shortcut minima.
These objectives regularize different parts of the system and can be jointly optimized.

\paragraph{Train-only regularization.}
Decoding heads are attached only during training and removed at inference, yielding zero additional online latency compared to standard embedding-based retrieval.

\subsection{Action Alignment Objective}

Given the ground-truth next clicked item $i_t$, we rank its embedding $\mathbf{e}_{i_t}$ above negatives $\{\mathbf{e}_j\}_{j\in\mathcal{N}_t}$, where $\mathcal{N}_t$ contains exposed-but-unclicked items (hard negatives) and in-batch negatives. We use a pairwise cross-entropy objective:
\begin{equation}
\mathcal{L}_{\text{act}}
=
\sum_{t}\sum_{j\in\mathcal{N}_t}
-\log\sigma\!\left(\mathbf{h}_t^\top\mathbf{e}_{i_t}-\mathbf{h}_t^\top\mathbf{e}_j\right)
\end{equation}
This form matches logged ranking supervision with heterogeneous negative hardness and is empirically more stable than temperature-scaled InfoNCE in our setting.

\subsection{Knowledge Regularization: Multimodal Decodability}

\subsubsection{Task~1: History-Conditioned Semantic Generation}
Let $\mathbf{y}_{i_t}=[w_1,\ldots,w_L]$ denote the target item's text tokens. We decode the target text from the behavioral history:
\begin{equation}
\mathcal{L}_{\text{gen}}
=
-\sum_{l=1}^{L}\log p_\theta\!\left(w_l\mid w_{<l}, \mathcal{H}_t\right)
\end{equation}
This term anchors training to the LLM's native token-level objective, mitigating catastrophic forgetting and discouraging the continuous-token interface from discarding recoverable semantics.

\subsubsection{Task~2: Embedding-Conditioned Semantic Reconstruction}
Our main anti-collapse constraint requires the next-interest embedding to be semantically decodable:
\begin{equation}
\mathcal{L}_{\text{recon}}
=
-\sum_{l=1}^{L}\log p_\theta\!\left(w_l\mid w_{<l}, \mathbf{h}_t\right)
\end{equation}
By enforcing an explicit embedding bottleneck, shortcut solutions that treat continuous item embeddings as opaque identifiers become suboptimal.

\subsubsection{Visual Decodability as a Shared Modality Decoder}
When multimodal supervision is available, we reconstruct a frozen visual semantic feature $\mathbf{v}_{i_t}$ (from a pretrained vision encoder) with a conditional diffusion/flow-matching head $g_\psi$. The head can be conditioned on either history states (Task~1) or $\mathbf{h}_t$ (Task~2). We optimize a standard diffusion/flow-matching training loss:
\begin{equation}
\mathcal{L}_{\text{img}}
=
\mathbb{E}_{\tau}\big[\ell_{\text{diff}}(g_\psi;\mathbf{v}_{i_t},\mathbf{c}_t,\tau)\big]
\end{equation}
where $\mathbf{c}_t$ denotes the conditioning signal and $\ell_{\text{diff}}$ follows standard formulations.

\subsubsection{Final Decodability Regularizer}
We instantiate the decodability regularizer as
\begin{equation}
\mathcal{L}_{\text{dec}}=\mathcal{L}_{\text{gen}}+\mathcal{L}_{\text{recon}}
\qquad
\mathcal{L}_{\text{dec}}^{\text{mm}}=\mathcal{L}_{\text{gen}}+\mathcal{L}_{\text{recon}}+\lambda_{\text{img}}\mathcal{L}_{\text{img}}
\end{equation}

\subsection{Bridging the Modality Gap with Staged Alignment}

Continuous item embeddings are out-of-distribution relative to the LLM's discrete-token pretraining. We therefore adopt a two-stage schedule: (i) a semantic warm-up that teaches the decoder to decode item semantics from isolated $\mathbf{e}_i$, aligning the continuous-token interface to the LLM space; and (ii) joint training on behavioral sequences with $\mathcal{L}_{\text{KARMA}}$, where action alignment learns personalization and decodability prevents collapse.

\subsection{Inference: Train-Heavy, Infer-Light}

At inference, KARMA uses only the efficient embedding path: encode history items into $\mathbf{e}$ tokens, run the causal decoder to obtain $\mathbf{h}_t$, and score candidates by dot-product similarity with item embeddings. All decoding heads are disabled, incurring zero additional online latency.

\section{Experiments}
\label{sec:experiments}

We evaluate KARMA in \textbf{Taobao Search}, one of the largest e-commerce search engines. The experiments are organized around three questions: (i) whether action-only training under continuous tokens exhibits semantic collapse, (ii) whether train-only decodability regularization mitigates collapse and improves retrieval/ranking, and (iii) how multimodal reconstruction and diffusion-based embedding generation behave under retrieval requirements. We additionally report online A/B results to validate practical impact under strict latency constraints.

\subsection{Experimental Setup}
\label{sec:setup}
\textbf{Data.} We use anonymized search logs with chronological splits. Each instance contains a user history of positive interactions, the next clicked item, and exposed-but-unclicked items as hard negatives.

\noindent\textbf{Metrics.} We report action metrics (gAUC for ranking discrimination; HR@K for retrieval/pre-ranking) and a lightweight semantic probe JS@50 (term-level Jaccard overlap between the ground-truth item and retrieved top-$K$ items), used to detect shortcut learning when action metrics remain competitive.

\noindent\textbf{Implementation.} Unless specified, we use Qwen3-0.6B \cite{yang2025qwen3} as the user decoder with the continuous-token interface; item encoders are Qwen3-1.7B (text) or Qwen3-VL-2B (multimodal) \cite{bai2025qwen3}. Decoding/reconstruction heads are train-only and removed at inference.

\subsection{Main Results: Decodability Regularization Mitigates Semantic Collapse}
\label{sec:main}

Table~\ref{tab:main} summarizes the effect of KARMA components, reported as improvements over the action-only baseline trained with Pairwise-CE. We highlight two observations.

\textbf{(1) Action-only training is prone to shortcut learning.}
Under the continuous-token interface, optimizing only $\mathcal{L}_{\text{act}}$ yields reasonable action metrics but comparatively weak semantic fidelity (JS@50), consistent with the model relying on non-semantic shortcuts rather than retaining decodable semantics in the interest embedding.

\textbf{(2) Embedding-conditioned decodability is the primary anti-collapse constraint.}
History-conditioned semantic generation (Task~1) provides a modest action gain but does not directly constrain the retrieval bottleneck $\mathbf{h}_t$. In contrast, embedding-conditioned reconstruction (Task~2) enforces that $\mathbf{h}_t$ remains semantically recoverable, leading to substantially larger retrieval improvements. Combining Task~1 and Task~2 yields the best overall performance, suggesting they regularize complementary failure modes.

\begin{table}[t]
\centering
\caption{Impact of decodability regularization ($\Delta$ over action-only Pairwise-CE).}
\label{tab:main}
\resizebox{\columnwidth}{!}{
\begin{tabular}{lcccc}
\toprule
\textbf{Variant} & \textbf{$\Delta$gAUC} & \textbf{$\Delta$HR@200} & \textbf{$\Delta$HR@1000} & \textbf{$\Delta$JS@50} \\
\midrule
\multicolumn{5}{l}{\textit{Text-only decodability}} \\
\quad + Task~1: History$\rightarrow$Text generation ($\mathcal{L}_{\text{gen}}$) & +0.43 & +4.81 & +3.46 & -0.87 \\
\quad + Task~2: $\mathbf{h}_t\rightarrow$Text reconstruction ($\mathcal{L}_{\text{recon}}$) & +0.43 & +19.19 & +19.31 & +2.65 \\
\quad \textbf{KARMA: Task~1 + Task~2} & \textbf{+0.97} & \textbf{+22.57} & \textbf{+21.19} & +2.26 \\
\midrule
\multicolumn{5}{l}{\textit{Multimodal extension (incremental over the best text-only KARMA)}} \\
\quad + Visual semantic reconstruction & \textbf{+1.38} & \textbf{+10.83} & \textbf{+8.41} & +0.40 \\
\bottomrule
\end{tabular}
}
\end{table}

\subsection{Diagnosing Collapse: Attention Sinks as Qualitative Evidence}
\label{sec:collapse}

\begin{figure}[t]
    \centering
    \begin{subfigure}[t]{0.49\columnwidth}
        \centering
        \includegraphics[width=\columnwidth]{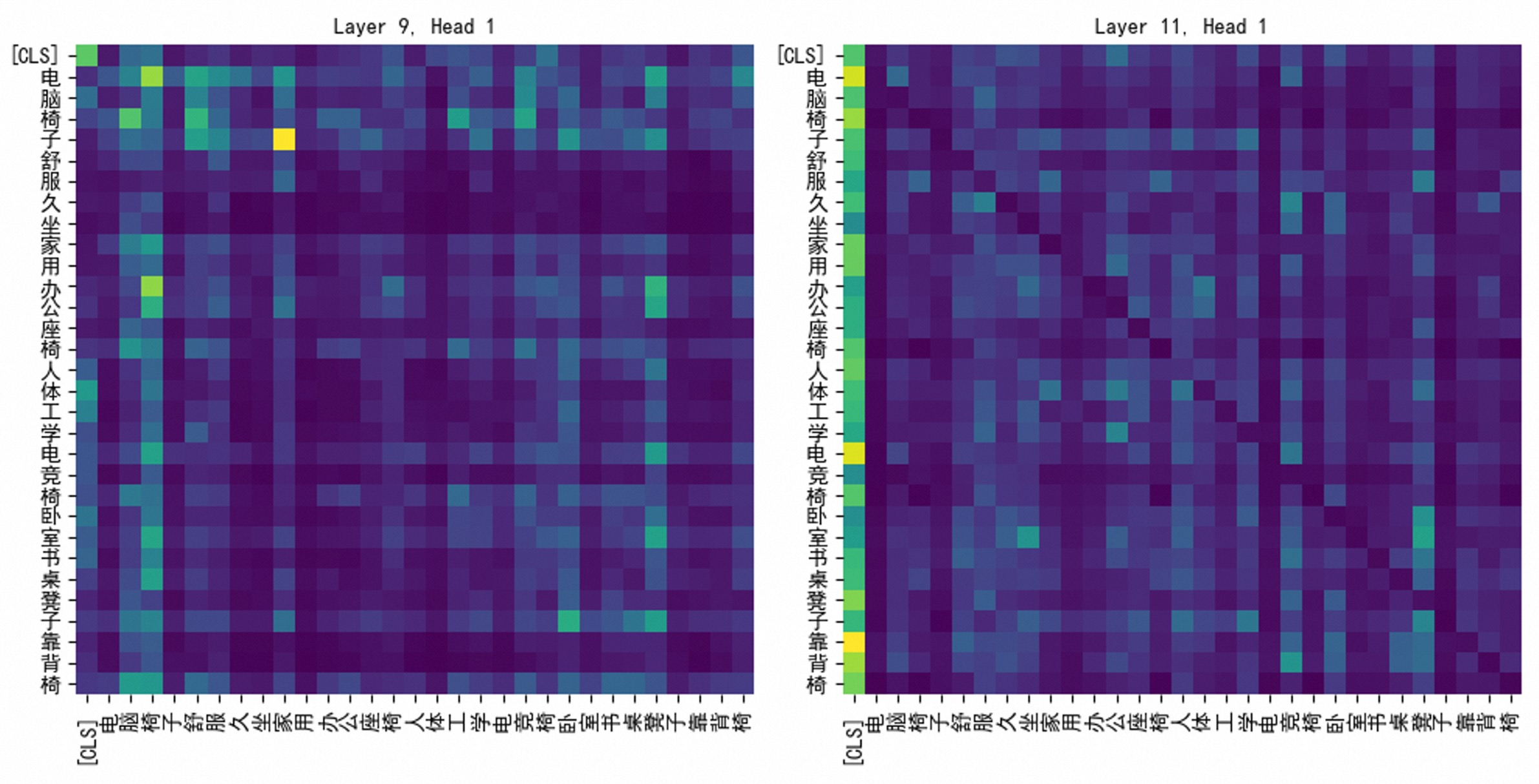}
        \caption{Frozen pretrained item encoder (no task training).}
    \end{subfigure}
    \hfill 
    \begin{subfigure}[t]{0.49\columnwidth}
        \centering
        \includegraphics[width=\columnwidth]{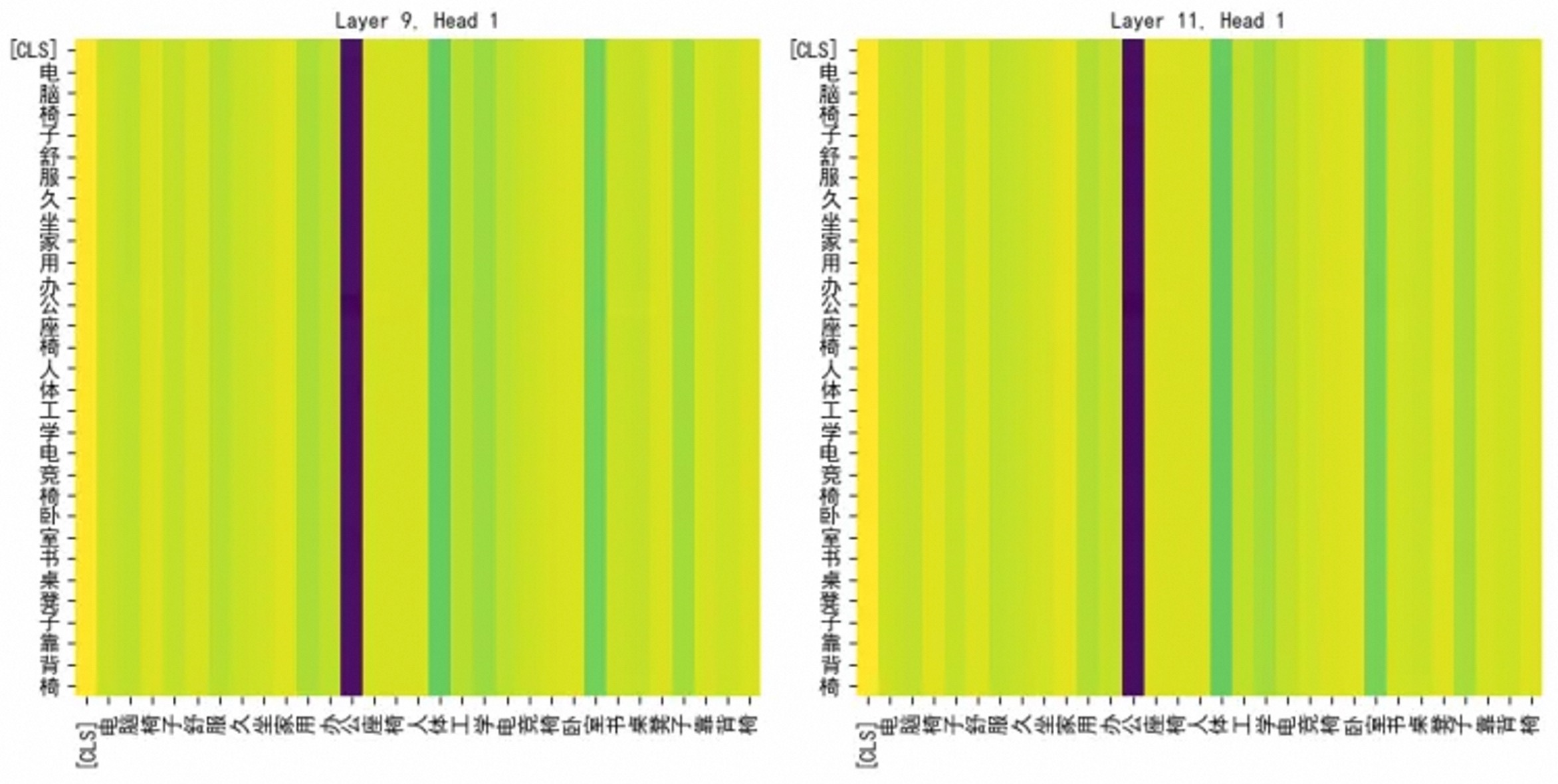}
        \caption{Action-only training with $\mathcal{L}_{\text{act}}$ exhibits attention sinks (barcode-like patterns).}
    \end{subfigure}

    \vspace{0em} 

    \begin{subfigure}[b]{0.49\columnwidth}
        \centering
        \includegraphics[width=\columnwidth]{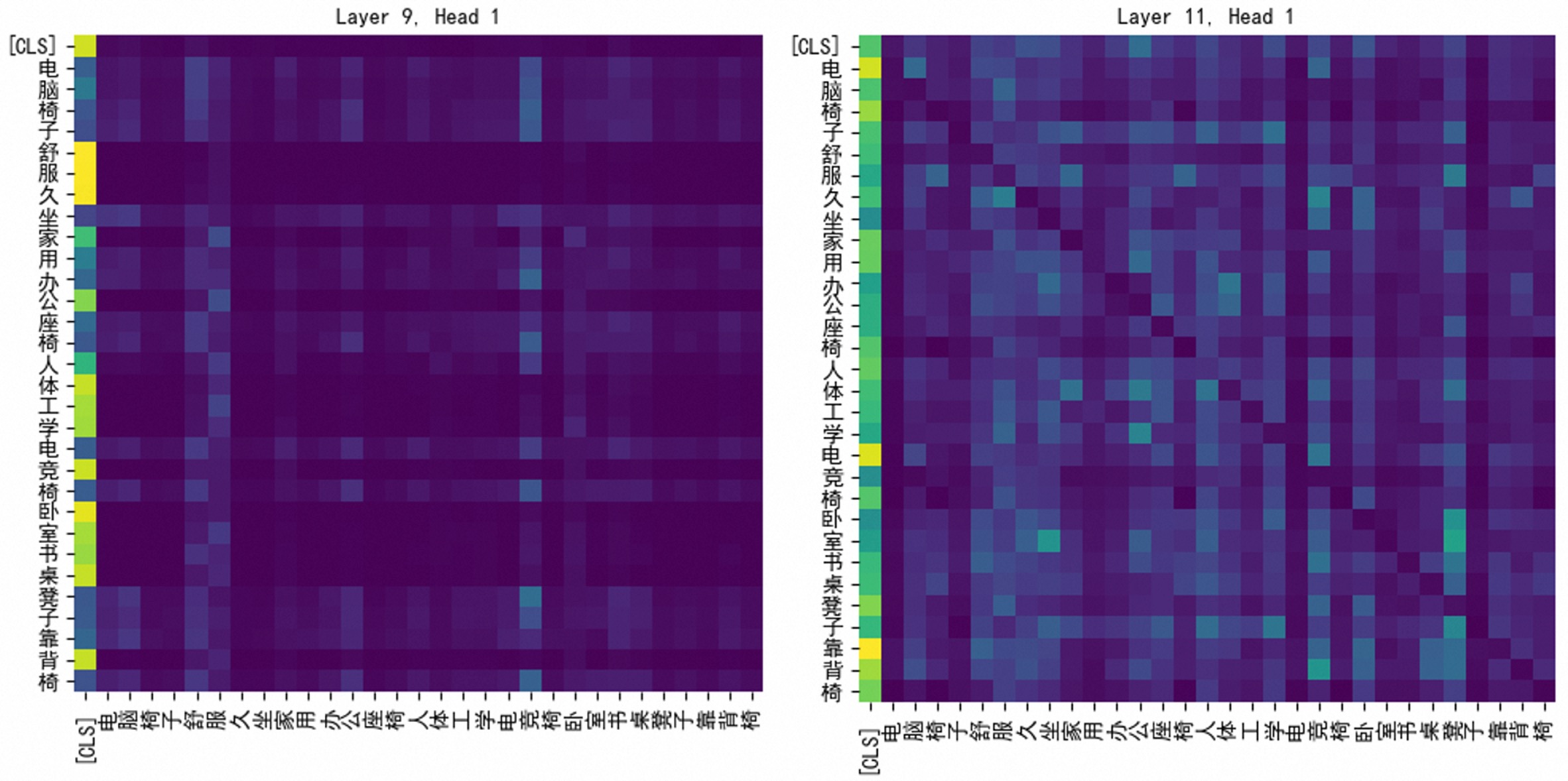}
        \caption{KARMA reduces attention sinks and restores more distributed attention.}
    \end{subfigure}
    



    \caption{Attention maps from certain layers (Layers 9/11, Head 1) of the item encoder under three training regimes.}
    \label{fig:attention}
\end{figure}

To better understand the failure mode, we visualize attention maps in the item encoder. Figure~\ref{fig:attention} shows that action-only training often produces sparse, barcode-like attention patterns that concentrate mass on a small subset of positions. This \emph{attention sink} behavior is consistent with shortcut learning under discriminative supervision: the encoder may over-focus on a few highly predictive cues while ignoring broader semantic content. With KARMA, attention becomes more distributed, suggesting that decodability regularization encourages representations that remain grounded in recoverable item semantics.

\paragraph{Note on interpretation.}
We treat attention sinks as a diagnostic signal rather than a causal proof. The main quantitative evidence remains the consistent gains in Table~\ref{tab:main} across action metrics and semantic fidelity proxies, together with the fact that all regularizers are removed at inference (zero online overhead).

\subsection{Multimodal Grounding and the Mode--Mean Dilemma}
\label{sec:mm}

\textbf{Multimodal grounding.}
When multimodal supervision is available, we add a lightweight diffusion/flow-matching head to reconstruct frozen visual semantic features of the target item. As shown in Table~\ref{tab:main}, visual reconstruction provides additional gains beyond text-only KARMA, indicating that cross-modal decodability offers complementary grounding signals for personalization.

\noindent\textbf{Negative result: diffusion is a poor generator for retrieval embeddings.}
We further tested whether diffusion can replace discriminative objectives to generate the retrieval embedding $\mathbf{h}_t$. Table~\ref{tab:diffusion} reports results \emph{relative to the best text-only KARMA}. Diffusion-based embedding generation underperforms, while a simple AR+MSE objective performs better.

\noindent\textbf{Insight: mode-seeking vs.\ mean-seeking.}
This discrepancy reflects a boundary between generative modeling and retrieval representation learning. Diffusion objectives are naturally \emph{mode-seeking}---they aim to sample high-fidelity instances from a conditional distribution. In contrast, retrieval embeddings are \emph{mean-seeking}---they should act as a stable centroid that aggregates multiple plausible future intents to support nearest-neighbor retrieval. Consequently, diffusion is effective as a semantic reconstruction regularizer, but suboptimal as the generator of retrieval embeddings.

\begin{table}[t]
\centering
\caption{Generating retrieval embeddings with diffusion ($\Delta$ over the best text-only KARMA).}
\label{tab:diffusion}
\resizebox{\columnwidth}{!}{
\begin{tabular}{lcccc}
\toprule
\textbf{Embedding Generator} & \textbf{$\Delta$gAUC} & \textbf{$\Delta$HR@200} & \textbf{$\Delta$HR@1000} & \textbf{$\Delta$JS@50} \\
\midrule
\textbf{AR + MSE} & \textbf{+0.63} & \textbf{+4.21} & \textbf{+1.70} & \textbf{+0.85} \\
AR + DDPM\cite{ho2020denoising} & +0.17 & +0.80 & -1.01 & +0.37 \\
AR + EDM\cite{karras2022elucidating} & -0.12 & -2.95 & -3.30 & -0.33 \\
AR + Flow-Matching\cite{lipman2022flow} & +0.02 & -0.83 & -2.68 & +0.26 \\
\bottomrule
\end{tabular}
}
\end{table}

\subsection{Online Deployment: Funnel-Wide Gains with Zero Inference Overhead}
\label{sec:online}
We applied KARMA across three stages of the system: ranking, pre-ranking, and recalling. The offline performance is presented as follows:
    
    \textbf{Ranking:} Compute the similarity between the user’s historical items and the target item as an input feature for the CTR model. +0.25 AUC.
    
    \textbf{Pre-ranking:} Weighted fusion of pre-ranking scores and KARMA scores. +1.86 HR@500.
    
    \textbf{Recalling:} Serving as an embedding-based retrieval. +2.51 HR @5000.
    

We deployed KARMA in production. Since all decodability heads are used only during training, the online serving path is identical to the action-only baseline. Over a 14-day A/B test at ranking \& pre-ranking stage, KARMA drives +0.9\% increase in GMV.

\section{Conclusions}

In this paper, we show that action-only discriminative training can trigger Semantic Collapse (e.g., attention sinks) and degrade semantic fidelity, limiting generalization. KARMA enforces \emph{train-only} semantic decodability via history-conditioned generation and embedding-conditioned reconstruction (optionally with visual feature reconstruction), preventing ID-like shortcut solutions while keeping the inference path unchanged. Across large-scale Taobao offline experiments and online A/B tests, KARMA consistently improves retrieval and ranking with zero serving overhead; we also find diffusion is better used as a multimodal semantic regularizer than as a mean-seeking retrieval-embedding generator.

\bibliographystyle{ACM-Reference-Format}
\bibliography{reference}

\end{document}